\documentclass{article}

\usepackage[utf8]{inputenc}
\usepackage[T1]{fontenc}
\usepackage{tabularx}
\usepackage{amssymb}
\usepackage{amsmath}
\usepackage{amsfonts}
\usepackage{amsthm}

\usepackage{arxiv}

\usepackage{booktabs}
\usepackage{graphicx}
\usepackage{float}
\usepackage{nicefrac}
\usepackage{microtype}
\usepackage{url}
\usepackage{hyperref}

\graphicspath{{./}}
\title{pAI-Econ-claude: A Gated Human-in-the-Loop Multi-Agent Architecture for AI-Assisted Economic Theory Development}

\author{
  Chen Zhu \\
  China Agricultural University \\
  \texttt{zhuchen@cau.edu.cn} \\
  \And
  Xiaolu Wang \\
  China Agricultural University \\
  \texttt{wangxiaolu77@cau.edu.cn} \\
  \And
  Weilong Zhang \\
  University of Cambridge \\
  \texttt{wz301@cam.ac.uk} \\
}

\begin{document}
\maketitle

\begin{abstract}
In many social-science research tasks, such as economics, LLM-based agents must produce outputs for which no cheap, task-complete, machine-readable correctness signal exists. This creates a distinctive reliability problem for multi-agent systems: how should generation, critique, coordination, and human judgment be organized when no component can certify the final result? We address this problem through \textbf{pAI-Econ-claude}, a gated, human-in-the-loop multi-agent architecture for AI-assisted economic theory development. Agents coordinate through a shared workspace of inspectable intermediate records; specialized gates diagnose targeted failure modes and recommend loopbacks without certifying correctness; and human checkpoints retain authority over decisions that are costly to reverse. Three principles guide the design: failures should be intercepted while the relevant choices remain visible, a checker without an oracle may diagnose but may not certify, and human attention should be concentrated at points of highest irreversibility. We evaluate the architecture on five matched economic-theory tasks against an ungated baseline. Two evaluators blinded to configuration agreed on all five pairwise rankings, preferring the gated architecture in four tasks and the baseline in one. Mean failure severity fell from 1.58 to 1.16, while overall usefulness rose from 2.60 to 3.10. The largest observed gain occurred when a reality check rejected a false market-structure premise and a proof review prompted revision of a false welfare claim. The negative case shows that scaffolding can also compress an economically important mechanism too aggressively. The results support a bounded claim: gated oversight improves the auditability of AI-assisted economic theory without substituting for formal verification, and the allocation of irreversible human judgment is a more informative design variable than pure agent autonomy. The workflow is publicly available at \url{https://github.com/maxwell2732/pAI-Econ-claude}.
\end{abstract}



\section{Introduction}
\label{sec:introduction}

LLM-based agents are increasingly used across social-science research for literature search, hypothesis generation, coding, empirical analysis, and manuscript drafting \cite{lu2024ai,schmidgall2025agentlaboratory,gottweis2025aicoscientist}. Yet many social-scientific outputs, especially theoretical mechanisms, causal interpretations, and welfare claims, lack a cheap, task-complete, machine-readable correctness signal. Partial checks may verify code, algebraic derivations, numerical solutions, or predictions against observed data, but they cannot jointly certify institutional fit, behavioral assumptions, equilibrium selection, and normative interpretation. This creates a distinctive design problem for multi-agent systems: how should generation, critique, coordination, and human judgment be organized when no component can certify the final result?

We study this problem in economic theory development, a demanding social-science setting in which an empirically relevant and formally coherent model may still rest on an inappropriate canonical lineage, restrictive assumptions, hidden proof gaps, or unsupported welfare interpretations. Empirical economics has strong norms for data construction, identification, econometric execution, and computational reproducibility \cite{gentzkow2014code,christensen2018transparency}, and none of them settles the mechanism behind an estimated effect, the question a structural model is built to answer \cite{low2017structural}. Algebraic re-derivation, numerical analysis, and counterexample search can test parts of a model, but they do not jointly certify institutional fit, assumption validity, equilibrium selection, or welfare interpretation. A paper may identify an effect convincingly while leaving open which canonical model family fits the institution, which assumptions generate the result, whether the comparative statics are non-trivial, and whether the welfare interpretation follows. When a language model fills that gap, fluency conceals five recurring failures, catalogued in Section~\ref{subsec:failure-modes}, which are instances of confident unsupported generation \cite{ji2023survey} arising at the points where domain judgment matters most.

Existing agentic research systems demonstrate the value of decomposing scientific work into specialized roles and persistent intermediate outputs. Systems for automated science and machine-learning research coordinate agents for ideation, experimentation, coding, review, and writing \cite{lu2024ai,schmidgall2025agentlaboratory,ifargan2024datatopaper}, and comparable multi-agent pipelines with in-process quality control serve biomedical research \cite{luo2025intention}. Role-specialized LLM organizations \cite{hong2024metagpt,qian2024chatdev,li2023camel,guo2024llmmas}, general orchestration frameworks such as AutoGen \cite{wu2024autogen}, and reasoning--acting paradigms such as ReAct \cite{yao2022react} establish related patterns beyond any single domain. Self-critique loops also improve output when a model can meaningfully inspect its own intermediate steps \cite{madaan2023self}. In companion work we develop a related pipeline for empirical economic research and argue that human oversight is what makes AI-assisted social science reliable \cite{zhu2026hler,zhu2026humanattention}, and related workflows organized around persistent shared files reduce the steering needed to turn a question into a structured workspace \cite{abdelmoneum2026paimsc}. Many of these systems operate in settings with executable code, experimental feedback, or benchmarked outputs. Those signals do not cover the full economic-theory task, which additionally requires institutional reality checking, canonical-lineage selection, an explicit equilibrium concept, assumption-sensitive welfare accounting, and a record of which claims were proved, illustrated, weakened, or left open.

We address this gap with a human-in-the-loop multi-agent workflow for constructing the theory component of an empirical economics study. The workflow decomposes modeling into named and inspectable stages running from puzzle refinement to manuscript construction, each writing a persistent record (Table~\ref{tab:pipeline}). Quality gates target specific failure modes and recommend loopbacks without certifying any stage as correct, and mandatory checkpoints reserve the institutional framing, the equilibrium concept, the proposition set, and the treatment of counterexamples for the researcher. Human oversight is therefore part of the system architecture \cite{mosqueira2023hitl,takerngsaksiri2025human}.

The central research question is comparative: holding the user-level task and underlying model fixed within each matched pair, does a gated oversight architecture produce more reliable and useful theory manuscripts than an ordinary ungated interaction? We evaluate the workflow using five matched task pairs spanning human capital, health economics, nutrition and macro-development, consumer information, and agri-food system transformation. Two evaluators scored the ten resulting manuscripts under blinded A/B labels and agreed on all five pairwise rankings, preferring the full workflow in four tasks and the baseline in one (Section~\ref{sec:evaluation}). The largest difference arose in the nutrition-label task, whose process trace recorded rejection of a false monopoly premise and revision of a false welfare-incidence proposition; Task 1 is the necessary counterexample, where both evaluators preferred the richer baseline because the full workflow simplified the human-capital mechanism without resolving its key formal errors. The stage-by-stage record shows why: a gate sometimes initiated a loopback that changed the model or proposition set, and elsewhere exposed a gap without closing it, so the architecture's value depends on whether the relevant check is well targeted and the human adjudication produces a valid correction.

This paper makes three contributions. First, it derives three architectural principles for agent systems whose outputs cannot be certified by a task-complete automated verifier and realizes them as a gated multi-agent architecture with blackboard coordination, a supervisory gate tier, and bounded agent autonomy. Second, it develops a failure taxonomy and maps each mode to a concrete interception mechanism, making reliability claims inspectable at the level of individual stage outputs. Third, it provides a matched evaluation protocol combining blinded endpoint ratings with process tracing over the workflow's own records. The broader implication is that, absent a reliable task-complete oracle, autonomy is a poor proxy for reliability, while the placement of irreversible human judgment, independent re-derivation, adversarial counterexample search, and disclosure of unresolved gaps are more informative design variables.


\section{Design Motivation}
\label{sec:motivation}

This section organizes the errors of LLM-assisted economic theory into a small taxonomy of failure modes, including citation hallucination and unreliable reference grounding, and then states three architectural principles that the taxonomy forces on any oversight system built to intercept them. The principles govern where stages, quality gates, and human checkpoints are placed in Sections~\ref{sec:system}--\ref{sec:model-library}, and Section~\ref{sec:evaluation} tests whether the resulting placement does the work claimed for it. The organizing premise is that reliability is carried by the oversight architecture, with the underlying model held fixed within each matched task pair.

\subsection{A Taxonomy of Failure Modes}
\label{subsec:failure-modes}

When a capable model turns an empirical puzzle into theory, its mistakes cluster into recognizable categories, each attached to a specific point in the modeling process and each requiring a different check to catch. Table~\ref{tab:failure-taxonomy} lists five.

\begin{table}[t]
\centering
\footnotesize
\begin{tabularx}{\textwidth}{@{}p{0.05\textwidth} p{0.19\textwidth} X@{}}
\toprule
ID & Failure mode & Description \\
\midrule
F1 & Canonical mismatch & Idea placed in the wrong model family, or a standard model relabeled as new; the natural failure of free-form generation, which has no incentive to locate a tradition. \\
F2 & Trivial propositions & A result follows directly from an assumption (e.g.\ a hard-wired monotonicity); the formal apparatus is real but empty. \\
F3 & Hidden proof gaps & A derivation skips cases, assumes a sign, or conflates sufficient and necessary conditions; such failures can survive ordinary reading and may require explicit obligation tracking, independent re-derivation, counterexample search, or numerical checking to expose. \\
F4 & Interpretive overreach & Economic reading exceeds the formal result: unsupported welfare claims, over-generalization, or asserted institutional facts. \\
F5 & Unreliable citations & Fabricated, misattributed, or unverified citations; especially damaging here because canonical lineage and institutional motivation are carried by references. \\
\bottomrule
\end{tabularx}
\caption{Failure taxonomy for LLM-assisted economic theory development. F1--F5 are concrete enough to be adjudicated against a finished manuscript and are scored in Section~\ref{sec:evaluation}; F5 is coded by checking bibliography usability and whether cited sources support the claims assigned to them.}
\label{tab:failure-taxonomy}
\end{table}

Each stage, gate, and human checkpoint is mapped to the failure mode it targets where the pipeline is introduced (Section~\ref{sec:system}, Table~\ref{tab:pipeline}), and Section~\ref{sec:evaluation} tests whether the gates catch what the taxonomy predicts. Three properties of Table~\ref{tab:failure-taxonomy} constrain how that mapping can be organized, and they are what the architecture of Section~\ref{sec:system} is built to satisfy.

\subsection{Three Architectural Principles}
\label{subsec:principles}

\textbf{P1. Interception must happen while the failure is still visible.} F1 and F3 leave no trace in the finished manuscript: a model placed in the wrong family still reads as a model, and a derivation that quietly assumes a sign still reads as a proof. Both are visible only while the choice is being made, when the rejected model families are still on the table and the discharged and undischarged proof obligations are still separable. An architecture emitting only a final manuscript therefore offers nothing to inspect. Stages that each write a persistent record are what create the inspection surface, which is why the pipeline is staged.

\textbf{P2. A checker without an oracle may diagnose and may not certify.} Every gate here is itself a language model, exposed to the failure modes it is asked to detect. A gate emitting a correctness certificate would move the risk of confident error from the stage agent to the gate agent and conceal it behind an apparent verification step. The implementation uses \textsc{pass}/\textsc{reframe} as operational labels, but \textsc{pass} means only that the gate did not detect its targeted failure under the stated checklist; it is not a certificate of correctness. A \textsc{reframe} verdict reports a failure reason, severity, recommended loopback stage, and whether an override is permitted, with adjudication left to the researcher. Certification belongs to domains possessing a task-complete external verifier; in its absence the achievable standard is auditability.

\textbf{P3. Human judgment belongs at the points of highest irreversibility.} Checkpoints consume researcher attention, the scarce resource in a human-in-the-loop system, so they cannot follow every stage. The ordering principle is the cost of a late reversal. The equilibrium concept fixes the meaning of every subsequent proposition, proof, and welfare statement, so an error there invalidates everything downstream and admits no local patch, whereas an over-stated interpretation can be corrected in place at any time. Hence HiL-4 is the one unconditional hard stop tied to a substantive modeling choice, with the remaining scheduled checkpoints placed where a decision constrains later work without wholly determining it. Negative gate verdicts create additional conditional pauses for human adjudication. The principle survives translation out of economics: in any pipeline lacking a task-complete external verifier, the placement of irreversible human judgment varies independently of how autonomous the agents are.


\section{System Architecture}
\label{sec:system}

This section instantiates P1--P3 as a skill-based orchestration layer for human-in-the-loop theoretical modeling. P1 becomes a staged decomposition in which every stage writes an inspectable record and canonical matching precedes construction, so institutional reality and literature grounding are checked while the model family can still change. P2 becomes a gate design emitting diagnoses and loopback recommendations without certifying any stage. P3 becomes a checkpoint schedule concentrated on decisions that cannot be reversed cheaply, of which fixing the equilibrium concept is the extreme case.

The architecture is organized around the three-phase pipeline shown in Figure~\ref{fig:architecture}: grounding the research puzzle (phase (a)), constructing the model through canonical matching and assumption auditing (phase (b)), and stress-testing and interpreting the resulting propositions (phase (c)). An orchestration core selects the entry mode and routes execution through this pipeline; the knowledge and workflow assets described below, including the canonical model library, prompt templates, documentation, and examples, feed the pipeline directly and form no separate architectural layer. Each stage and gate box in Figure~\ref{fig:architecture} is executed by one specialized agent: a stage agent (dark-blue marker) produces the corresponding output, a gate or verdict agent (teal marker) evaluates it and emits an operational \textsc{pass}/\textsc{reframe} verdict, and the reality-check and counterexample-finder stages (red marker) are adversarial critique agents, charged with attacking work they did not produce; the persona council at Stage 3+3b further decomposes into four sub-agents, discussed in Section~\ref{subsec:mas}. Here \textsc{pass} denotes only that no targeted failure was detected. The purple HiL labels mark scheduled human checkpoints, while a negative gate verdict creates a conditional adjudication pause. The numbered gate labels in Figure~\ref{fig:architecture} are implementation identifiers; Table~\ref{tab:pipeline} uses functional names. Oversight sits above the agent roster as a separate layer, and it carries the reliability claim of Section~\ref{sec:motivation}.

\subsection{Entry Modes and Orchestration Core}

Economists arrive at theory from different starting points, so four entry modes share one core: \textit{Model Extension} adds a mechanism to a known canonical model; \textit{Phenomenon-to-Model} takes an empirical or institutional phenomenon and recommends model families; \textit{Model Critique} returns a referee-style audit of an existing draft; \textit{Full Pipeline} proceeds from intuition to manuscript skeleton. The orchestration core selects the mode and decomposes theory development into inspectable decisions: what is the puzzle, which model family applies, which institutional facts need checking, which assumptions do the work. That decomposition is what lets gates and checkpoints act on the process itself, at every intermediate step.

\subsection{Theory-Building Pipeline}

\begin{figure}[t]
 \centering
 \includegraphics[width=\textwidth]{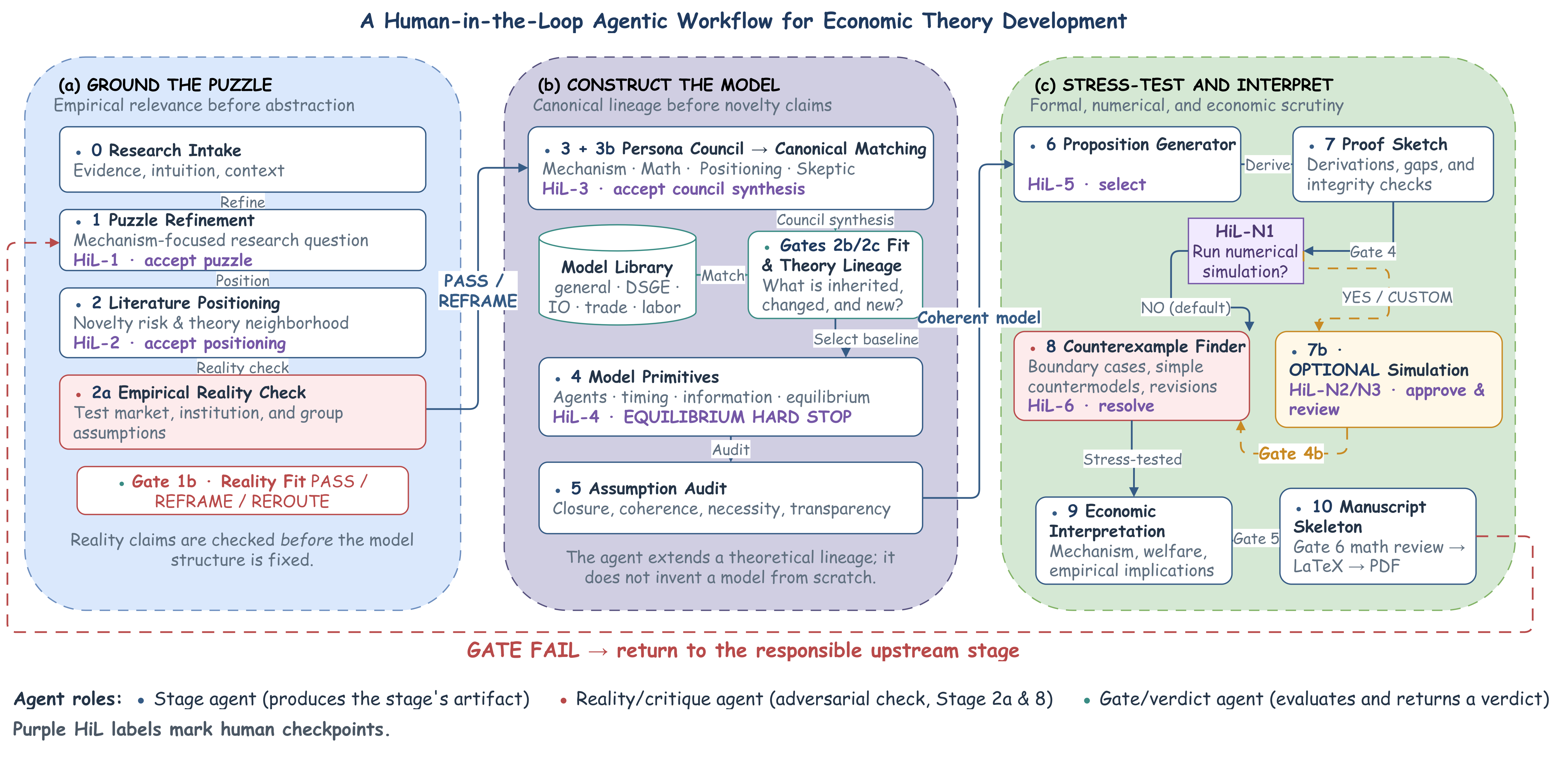}
 \caption{Architecture of the pAI-Econ-claude workflow, in three phases: grounding the research puzzle, constructing the model through canonical matching and assumption auditing, and stress-testing the resulting propositions through proof review, optional numerical simulation, counterexample search, and economic interpretation. Purple HiL labels mark scheduled human checkpoints. Gate verdicts can recommend revision, but the loopback arrows are executed only after human adjudication.}
 \label{fig:architecture}
\end{figure}

The pipeline is a Stage 0--10 sequence with three inserted sub-stages, 2a, 3b, and the optional numerical stage 7b, organized into the three phases of Figure~\ref{fig:architecture}. Each stage writes a named file that can be inspected, revised, and reused. Table~\ref{tab:pipeline} lists the stages together with the gate and human checkpoint that follow each, and the failure mode(s) from Table~\ref{tab:failure-taxonomy} each primarily targets.

\begin{table}[t]
\centering
\footnotesize
\begin{tabularx}{\textwidth}{@{}l X l l l@{}}
\toprule
Stage & Name & Gate after & HiL after & Targets \\
\midrule
0 & Intake & -- & -- & -- \\
1 & Puzzle Refinement & -- & HiL-1 & -- \\
2 & Literature Positioning & Novelty Risk & HiL-2 & F5 \\
2a & Empirical Reality Check & Reality Fit & -- & F1, F4 \\
3 & Theory Persona Council & -- & HiL-3 & F1 \\
3b & Canonical Model Matching & Canon.\ Fit, Lineage & -- & F1 \\
4 & Model Primitives & Model Coherence & HiL-4 (hard) & F1, F2 \\
5 & Assumption Audit & -- & -- & F2 \\
6 & Proposition Generator & Non-triviality & HiL-5 & F2 \\
7 & Proof Sketch & Proof Integrity & -- & F3 \\
7b & Numerical Analysis (optional) & -- & HiL-N1--N3 & F3 \\
8 & Counterexample Finder & -- & HiL-6 & F3 \\
9 & Economic Interpretation & Economic Meaning & -- & F4 \\
10 & Manuscript Skeleton & Math Review & -- & F3, F4 \\
\bottomrule
\end{tabularx}
\caption{The Stage 0--10 pipeline with inserted sub-stages 2a, 3b, and 7b. Each stage writes a named output file; gates and scheduled human checkpoints (HiL) attach after the stages indicated. HiL-4, fixing the equilibrium concept, is the only unconditional hard stop tied to a substantive modeling choice (P3); any negative gate verdict creates a separate conditional pause for adjudication. Stage 7b is entered only if the researcher approves a numerical plan at HiL-N1, with results approved and reviewed at HiL-N2 and HiL-N3. Targets lists the failure modes from Table~\ref{tab:failure-taxonomy} the stage or its gate primarily intercepts; Stages 0 and 1 are preparatory.}
\label{tab:pipeline}
\end{table}

Four stages carry most of the interception load. Stage 2a's reality check and Stage 3b's canonical matching act before the primitives are fixed, while Stage 7's proof sketch and Stage 8's counterexample search act afterward, once propositions exist to test. The remaining stages prepare or consolidate the material these checks read.

\subsection{Quality Gates, Human Oversight, and the Run Workspace}

The workflow includes nine gates: novelty risk, reality fit, canonical fit, theory lineage, model coherence, non-triviality, proof integrity, economic meaning, and a final math review at the manuscript stage. Following P2, no gate certifies correctness or loops silently. A \textsc{reframe} verdict pauses execution and reports a failure reason, severity, recommended loopback stage, and whether an override is permitted. The researcher may confirm the loopback, override it with a recorded justification, edit the output, or terminate the run. The six named HiL checkpoints from puzzle revision (HiL-1) through counterexample treatment (HiL-6) are unconditional scheduled reviews, with three further checkpoints (HiL-N1 to HiL-N3) governing entry into and use of the optional numerical stage. HiL-4 is the only unconditional hard stop tied to a substantive modeling choice, requiring the researcher to fix the equilibrium concept (competitive, Nash, subgame-perfect, Bayesian Nash, perfect Bayesian, or planner's solution) before the pipeline continues. Each run produces a workspace of staged outputs, gate reports, adjudication records, and a stage-level log. This workspace supports auditability where substantive correctness cannot be certified mechanically, complementing rather than replacing computational reproducibility.

\subsection{Relation to Multi-Agent Systems}
\label{subsec:mas}

The system is a multi-agent organization with a deliberately restricted autonomy budget, and P1--P3 constrain each of its multi-agent design choices. Operationally, an agent is a role-specific model invocation with a distinct prompt, task obligation, and read--write interface to the shared workspace. Agents may share the same underlying language model, but they remain organizationally separated by their roles, permitted actions, and positions in the control flow.

\textbf{Coordination is mediated by a shared workspace.} Agents never message one another directly. Each stage agent reads the files written by its predecessors and writes its own into a persistent workspace, which gate agents also read to produce verdicts. The coordination model is a blackboard: control passes through inspectable shared state, with no point-to-point channel between agents. P1 makes this load-bearing, since the shared state is exactly the inspection surface that direct messaging would leave transient, and it gives the human one place to intervene, a checkpoint being an edit to the blackboard between two agent activations. The overlap with the Agents and Artifacts meta-model \cite{omicini2008artifacts} is partly substantive: these workspace files are passive, persistent, agent-readable environment entities in the sense of an A\&A artifact, but they expose no operations for agents to invoke, placing them closer to a structured blackboard than to a full A\&A artifact.

\textbf{Organization is supervisory.} Gate agents stand in a normative relation to stage agents, evaluating a stage output against a stated obligation and emitting a verdict, a severity, and a recommended loopback, while holding no authority to enforce it. Enforcement rests with the human at either a scheduled checkpoint or a conditional gate-adjudication pause, where the researcher may accept a loopback, override it, edit the output, or terminate the run. P2 forces this separation of diagnosis from enforcement, yielding a three-tier organization in which the lower tier produces, the middle tier critiques, and the upper tier decides. The reliability claim of this paper concerns that division of authority. No claim is made about the competence of any single agent.

\textbf{Autonomy is bounded by construction.} Stage agents choose how to fill their assigned output and never whether to run; activation order is fixed by the orchestration core except at loopbacks, where a gate recommends the target stage and a human confirms it. This is far less autonomy than role-based LLM organizations that let agents negotiate task allocation \cite{hong2024metagpt,qian2024chatdev,li2023camel,guo2024llmmas}. The restriction is the point: without a task-complete external verifier, autonomy in task allocation adds failure modes while adding no signal that would reliably catch them, so the design spends its autonomy budget where disagreement is informative.

\textbf{The persona council is the one parallel-role structure.} Stage 3+3b activates four sub-agents on the same puzzle: a mechanism agent proposing the economic force at work, a formalization agent proposing how to write it down, a positioning agent proposing where it sits in the literature, and a skeptic charged with attacking all three. They return independent proposals, and the synthesis reaches the researcher at HiL-3 with the skeptic's objections attached and unresolved. Disagreement is surfaced to the human and left open inside the council, which separates the design from debate procedures that aggregate toward a single answer \cite{du2024debate}. P2 explains why: an internal aggregation rule is certification by majority, which a system without an oracle cannot justify.


\section{The Canonical Model Library and Theory-Lineage Protocol}
\label{sec:model-library}

The canonical model library, with the protocol that uses it to enforce theory lineage, is what most distinguishes this workflow from a general research agent.

\subsection{Library Structure and Theory-Lineage Protocol}

The library holds structured model templates: each entry records, for one model family, when to use it, its primitives and timing, information structure and equilibrium concept, typical comparative statics and extensions, and its common failure modes. A general layer of sixteen entries covers consumer and discrete choice, search, costly information acquisition, signaling, screening, moral hazard, adverse selection, product differentiation, mechanism design, matching, dynamic optimization, overlapping generations, principal-agent, general equilibrium, and political economy; a human-capital sub-library adds Becker, Ben-Porath, Roy, Cunha-Heckman skill formation, education under credit constraints, and the Acemoglu-Restrepo task framework. One restriction is deliberate: the library admits only \emph{structural} models carrying an equilibrium concept and provable propositions, which excludes econometric identification frameworks and keeps matching focused on economic structure.

The library feeds canonical matching (Stage 3b), enforced through two gates: the canonical-fit gate asks whether the proposed model is genuinely matched to its claimed family or a renamed instance of another, and the theory-lineage gate requires an explicit statement of what the model inherits, modifies, and adds relative to its canonical ancestor, together with a demonstration that the new result does not follow from the unmodified ancestor. Both target F1 and force a contribution to be expressed as a delta against a named ancestor, which disciplines novelty claims and makes the manuscript legible to referees reading it against the same tradition.


\subsection{Two Demonstrations of Canonical Matching}
\label{sec:demos}

Two runs illustrate the protocol in different entry modes. In \textit{Model Extension Mode}, an extension of Becker's human capital model to China's urban-rural divide was placed in the Becker tradition with a genuine gap flagged (no group-specific return function) and a second family surfaced (credit constraints). In \textit{Phenomenon-to-Model Mode}, a carbonated-drink market converged on third-degree price discrimination with dominant-firm pricing and a cross-market capacity constraint, while a Hotelling framing was rejected (the urban-rural gap traces to demand elasticity, with spatial preference judged immaterial) and screening set aside (observable types). The rejections are the instructive part: canonical matching intercepts F1 by ruling out plausible-looking families and recording the reason, a discipline free-form generation has no incentive to impose.


\section{Evaluation}
\label{sec:evaluation}

The demonstrations show that the workflow runs end to end without establishing that the checks improve reliability. The evaluation asks whether evaluators prefer the gated workflow's final manuscripts and, when a difference appears, whether the process record identifies the stage, gate, or human decision associated with the revision path.

\subsection{Matched-Task Design and External Evaluation}

Each task ran under two configurations on the same short user prompt, in a separate workspace and session with auto-memory disabled. The \emph{no-skill baseline} operated in a clean generic workspace with no access to the workflow repository, library, staged prompts, gates, or checkpoints, approximating a direct request to a general AI research assistant. The \emph{full-workflow configuration} operated in a clean copy of the workflow repository with the complete pipeline, and human intervention was admitted only at scheduled checkpoints or in response to negative gate verdicts, with every intervention recorded in the stage log. The five tasks covered human capital and hukou, centralized drug procurement, nutrition-capital DSGE modeling, nutrition-label industrial organization, and agri-food system transformation. Four were stated as short ordinary research requests; Task 4 deliberately embedded a false institutional premise to test whether the reality-check stage would intercept it before model construction. The underlying model was held fixed within each matched pair but differed across pairs: Sonnet 4.6 for Task 1, Sonnet 5 for Tasks 2 and 3, and Fable 5 for Tasks 4 and 5. Metered against the plan-relative rolling five-hour usage allowance, the full workflow consumed 4.6 to 18 times the allowance used by its paired baseline. Because this measure is not a token count and model versions draw on it at different rates, it is interpretable only within pairs; the present evaluation does not establish that the workflow is cost-effective.

The ten resulting manuscripts were relabeled A/B within task, and two evaluators, one model-based and one a human economist who took no part in designing the workflow, selecting the tasks, or running either configuration, completed the same structured scoring form without access to configuration labels or intermediate outputs, recording independent scores plus a pairwise preference, preference magnitude, and confidence. Failure severity is scored 0--3 (0 = not present, 3 = severe, lower preferred) along F1--F5; positive usefulness dimensions are scored 1--5 (higher preferred): canonical-model fit, mechanism originality, internal consistency, proof integrity, reality correspondence, testable predictions, literature reliability, writing and organization, and a holistic overall-usefulness rating. After blinded scoring, configurations were unblinded and the intermediate outputs were separately reviewed and classified as flagged, resolved, partially resolved, or missed relative to each external finding; the external scores are the primary endpoint measures, and the review of the workflow's own records supplies process evidence.

The evaluators agreed on all five pairwise rankings, preferring the full workflow in four tasks and the baseline in one; the mean preference magnitude was 1.60 on a 0--3 scale, with the largest gap in Task 4. Table~\ref{tab:external-task-summary} reports the task-level comparison.

\begin{table}[tp]
\centering
\footnotesize
\setlength{\tabcolsep}{4pt}
\begin{tabular}{@{}l l c cc cc c@{}}
\toprule
& & & \multicolumn{2}{c}{Failure} & \multicolumn{2}{c}{Overall} & \\
\cmidrule(lr){4-5}\cmidrule(lr){6-7}
Task & Preferred & Gap & Full & Base & Full & Base & Diff.\\
\midrule
T1 Human capital & Base & 1.50 & 1.80 & 1.40 & 2.50 & 3.00 & $-0.50$ \\
T2 Drug procurement & Full & 1.00 & 1.20 & 1.80 & 3.00 & 2.50 & $+0.50$ \\
T3 Nutrition DSGE & Full & 1.50 & 1.10 & 1.50 & 3.00 & 2.50 & $+0.50$ \\
T4 Nutrition label & Full & 2.50 & 0.70 & 1.40 & 4.00 & 2.50 & $+1.50$ \\
T5 Agri-food DSGE & Full & 1.50 & 1.00 & 1.80 & 3.00 & 2.50 & $+0.50$ \\
\midrule
Mean (Full in 4/5) & -- & 1.60 & 1.16 & 1.58 & 3.10 & 2.60 & $+0.50$ \\
\bottomrule
\end{tabular}
\caption{Task-level results. Failure is the mean of F1--F5 across evaluators (lower better); Overall is the mean holistic-usefulness rating (higher better); Gap is the mean pairwise preference magnitude on a 0--3 scale; Diff.\ is full minus baseline Overall.}
\label{tab:external-task-summary}
\end{table}

Descriptively averaged over the five heterogeneous task--model pairs and two evaluators, the full workflow is associated with 0.42 lower mean failure severity and 0.50 higher overall usefulness (Table~\ref{tab:external-dimension-summary}). The largest failure differences concern interpretive overreach, assumption-driven propositions, and proof gaps; the largest usefulness differences concern internal consistency, proof integrity, and writing. The baseline scores slightly higher on reality correspondence and testable predictions, since its broader models sometimes preserve more institutional detail, while literature reliability ties. The pattern suggests a trade: the full workflow gains 1.10 on internal consistency and 0.80 on proof integrity but loses 0.10 on reality correspondence and 0.30 on testable predictions. Task 1 displays the same tension in sharper form, which is why every dimension is reported separately.

\begin{table}[tp]
\centering
\footnotesize
\begin{tabular}{@{}lccc@{}}
\toprule
Dimension & Full workflow & Baseline & Full minus baseline \\
\midrule
\multicolumn{4}{l}{\emph{Panel A: Failure severity, lower is better}} \\
F1 Canonical mismatch & 0.80 & 0.90 & $-0.10$ \\
F2 Trivial / assumption-driven & 1.10 & 1.60 & $-0.50$ \\
F3 Proof gaps & 2.00 & 2.50 & $-0.50$ \\
F4 Interpretive overreach & 1.10 & 1.90 & $-0.80$ \\
F5 Citation grounding & 0.80 & 1.00 & $-0.20$ \\
Mean failure severity & 1.16 & 1.58 & $-0.42$ \\
\midrule
\multicolumn{4}{l}{\emph{Panel B: Positive usefulness, higher is better}} \\
CMF Canonical-model fit & 3.60 & 3.40 & $+0.20$ \\
MO Mechanism originality & 3.00 & 2.70 & $+0.30$ \\
MIC Internal consistency & 3.30 & 2.20 & $+1.10$ \\
PI Proof integrity & 2.50 & 1.70 & $+0.80$ \\
RC Reality correspondence & 3.10 & 3.20 & $-0.10$ \\
TPI Testable predictions & 3.20 & 3.50 & $-0.30$ \\
LRP Literature reliability & 3.40 & 3.40 & $0.00$ \\
WO Writing and organization & 4.00 & 3.40 & $+0.60$ \\
Overall usefulness & 3.10 & 2.60 & $+0.50$ \\
\bottomrule
\end{tabular}
\caption{Descriptive dimension-level means across five heterogeneous task--model pairs and two external evaluators. Negative differences are favorable in Panel A; positive differences are favorable in Panel B.}
\label{tab:external-dimension-summary}
\end{table}

Agreement on pairwise ranking exceeds agreement on exact scores: the evaluators differed in severity on several dimensions, and one repeatedly noted scope asymmetries between complete papers and shorter theory-section excerpts. The numerical differences are therefore descriptive, and the unanimous pairwise ordering is the more stable result.

\subsection{Task 1: A Negative Case in Human-Capital Modeling}

Both evaluators preferred the baseline (difference magnitude 1.50): the full workflow scored higher failure severity (1.80 versus 1.40) and lower overall usefulness (2.50 versus 3.00). The baseline preserved a broader institutional mechanism spanning school quality, credit constraints, migration barriers, and hukou-related labor-market access, which both judged more economically meaningful despite unresolved migration-threshold cases. The full workflow compressed the problem into a cleaner binary-threshold model without a matching formal gain, and external review found an internally inconsistent policy-switch threshold and an under-conditioned welfare-supermodularity claim. The stage record shows canonical matching, novelty screening, and reference verification all ran and flagged the relevant risk assumptions, while the decisive errors survived. More structure can therefore narrow a model prematurely, and a gate can create confidence without removing the decisive error. Task 1 also reverses the initial author-side preference for the more polished manuscript, which is why endpoint evaluation must stay separate from process self-assessment. Because model versions differed across task pairs, variation in the observed workflow advantage across Tasks 1--5 cannot be attributed to task characteristics alone; the within-Task-1 comparison itself nevertheless holds the model fixed.

\subsection{Tasks 2, 3, and 5: Detection Without Full Resolution}

Both evaluators preferred the full workflow in the three remaining tasks (Table~\ref{tab:external-task-summary}; overall usefulness 3.00 versus 2.50 in each), all following a detect-and-narrow pattern short of a solved theorem. In Task 2 (drug procurement), the baseline's welfare accounting risked double-counting and its deadweight-loss formula was wrong; the proof gate rejected a uniqueness step and a counterexample search falsified a universal cost-dispersion claim (CRRA with uniform costs gives a corner, linear demand an interior peak), narrowing the claim to two regimes while leaving the primitive condition separating them open. In Task 3 (nutrition-capital DSGE), the baseline's capital-return, depreciation, and fiscal-block equations were mutually inconsistent; the numerical stage traced an apparent subsidy advantage to a non-optimal solver critical point, and constrained maximization reversed the ranking to transfer-dominance, leaving steady-state uniqueness open. In Task 5 (agri-food DSGE), the baseline ranked policy instruments by the largest wedge although each instrument moves several distorted quantities; a numerical sweep found 378 steady-state solve failures traced to a learning-loop gain, and the tightened domain was back-propagated to the propositions, leaving global determinacy open. In each case, the process record identifies a gate, numerical check, or counterexample search that exposed an error left silent in the baseline and prompted a narrower final claim. Proof integrity nevertheless remains the lowest usefulness dimension because the remaining obligations are disclosed rather than discharged.

\subsection{Task 4: Institutional Reframing and Proposition Repair}
\label{subsec:task4}

Task 4 produces the largest and most consistent observed gain: both evaluators preferred the full workflow (difference magnitude 2.50), failure severity fell from 1.40 to 0.70, and overall usefulness rose from 2.50 to 4.00. The prompt deliberately described the U.S. candy market as a monopoly, and the baseline accepted the premise and built a thin, parameter-dependent monopoly-attention model. The reality check classified the premise as false and issued a negative verdict, creating a conditional adjudication pause. The researcher confirmed the recommended reframing, rerouting the model to differentiated duopoly before primitives were fixed and producing a benchmark in which only one-third of the fixed-price healthy-share response survives strategic repricing. A second revision path was initiated by the proof gate: a proposition claiming that every informed switcher gains from label simplification was independently re-derived as false, returned upstream, and rewritten around an explicit attention-cost threshold. A subsequent numerical sweep showed that the result holds under uniform taste density and varies substantially outside it, so the manuscript reports the relative-price prediction with more confidence than the price-level prediction. Task 4 provides the clearest process evidence that targeted checks and human-confirmed loopbacks can alter final economic content rather than merely its presentation.

Taken together, the evaluation supports a bounded comparative claim. The largest observed gain occurs where the process trace records a targeted check followed by a valid human-confirmed revision (Task 4), while the gains are weaker where the pattern is detect-and-narrow (Tasks 2, 3, and 5). Task 1 prevents a universal conclusion: gates guarantee neither a correct proposition set nor the retention of useful institutional structure. The architecture is therefore best read as a reliability scaffold whose performance depends on gate coverage, human judgment, and the fit between the task and model library, rather than as evidence for the isolated causal effect of any single component.


\section{Limitations}
\label{sec:limitations}

This study has several limitations. First, the gates diagnose potential failures rather than certify correctness: proof sketches remain provisional, and numerical checks can reveal inconsistencies or counterexamples but cannot establish general theorems. Second, the evaluation covers only five task pairs and two evaluators, one model-based and one human economist acquainted with the authors; the reported score differences are therefore descriptive, and variation in output length and completeness may have affected the comparison. Third, the full workflow combines staged decomposition, a canonical model library, quality gates, human checkpoints, and numerical analysis, while also receiving more expert intervention than the baseline, so the contribution of any individual component cannot be isolated without ablation studies and matched human-attention budgets. Finally, the literature checks, institutional grounding, and model library have bounded coverage, and the implementation was tested with one commercial language-model family through one tooling environment, with model versions differing across task pairs. The system should therefore be treated as an auditable scaffold rather than a verification mechanism: novelty, factual accuracy, formal validity, and final interpretation remain the responsibility of the researcher, and broader expert evaluation and cross-model replication are needed.


\section{Conclusion}
\label{sec:conclusion}

This paper addresses a central design problem for AI-assisted social science: how to organize agentic workflows when the quality of the final output cannot be judged against a clear, task-complete external oracle. This problem is especially acute in social science because validity depends not only on formal consistency, but also on institutional context, behavioral assumptions, causal interpretation, and the boundary between positive analysis and normative judgment. Economic theory is a demanding case: a model may appear fluent and sophisticated while still containing canonical mismatches, assumption-driven propositions, proof gaps, or unsupported welfare claims. We therefore develop a gated human-in-the-loop architecture guided by three principles: intercept failures while key choices remain visible, let gates diagnose rather than certify, and place human judgment at decisions that are costly to reverse.

The matched-task evaluation provides qualified evidence that this design can improve AI-assisted theory development. The full workflow was preferred in four of five tasks, with the largest observed gain occurring where the process trace recorded an early diagnostic check followed by institutional reframing and proposition repair. The negative human-capital case, however, shows that additional structure can also compress a socially important mechanism too aggressively. The architecture's main contribution is therefore to make the modeling process more transparent and auditable by recording assumptions, rejected alternatives, unresolved obligations, and the points at which human judgment changed the analysis.

These features matter particularly for AI for social science. Unlike domains in which outputs can be checked against test accuracy, executable code, or formal proofs, many social-scientific claims remain contestable even after the calculations are correct. Competing models may fit the same empirical pattern, institutions may not travel across settings, distributional consequences may depend on unobserved heterogeneity, and welfare conclusions may embed value judgments that cannot be delegated to an automated checker. Reliability therefore depends on how evidence, theory, institutional knowledge, and human judgment are organized together. The relevant design question is not simply how much autonomy an agent should receive, but which decisions should remain reversible, which claims require independent challenge, and where human expertise has the greatest marginal value. A well-designed social-science agent should leave behind more than a polished answer: it should expose the reasoning structure, assumptions, alternatives, failed checks, revisions, and residual uncertainties that determine how much confidence the answer deserves. These broader implications are architectural hypotheses derived from the economic-theory case rather than empirically established results across social-science domains.

The architecture developed here provides a foundation for that broader agenda. Future work should isolate the contribution of individual components through ablation studies, equalize human-attention and output budgets across conditions, expand evaluation to larger and more diverse expert panels, and test the design across model families, language models, and social-science domains. Progress in AI for social science will depend on systems that can produce increasingly sophisticated analyses while preserving institutional sensitivity, interpretive restraint, and meaningful human accountability.


\section*{Code and Data Availability}

The implementation is released as an open-source Claude Code skill at
\url{https://github.com/maxwell2732/pAI-Econ-claude}, archived at
\url{https://doi.org/10.5281/zenodo.20686026} \cite{zhu2026paiecon}. The
repository contains the staged prompts, the canonical model library, the gate
specifications, and the evaluation artifacts for the five matched task pairs
reported in Section~\ref{sec:evaluation}.

\section*{Use of Generative AI in This Work}

Generative AI assisted manuscript preparation, including literature
organization, drafting and editing support, and \LaTeX{} formatting. All
experimental design, analysis, interpretation, and final judgments remain the
responsibility of the human authors, who take full responsibility for every
claim made. Generative AI is not an author.

\section*{Competing Interests}

The authors have no competing interests to declare that are relevant to the
content of this article.


\bibliographystyle{unsrt}
\bibliography{references_arXiv}


\appendix

\section*{Appendix}
\addcontentsline{toc}{section}{Appendix}

\noindent The appendices below provide the full task prompts, the evaluator-level
score tables, and the complete task-by-task process-tracing narratives
underlying the evaluation reported in Section~\ref{sec:evaluation}.

\section{Full Task Prompts}

Both configurations within a task received the identical short user-level research prompt described below; only the invocation environment differed (Section~\ref{sec:evaluation}). The Task 4 prompt is reproduced verbatim including the false institutional premise it embeds, namely that the U.S. candy market is a monopoly; Section~\ref{subsec:task4} reports how the reality-check stage handled it.

\paragraph{Task 1 (Human capital and hukou) baseline prompt.}
\begin{quote}\small
I am working on an empirical economics paper about China's urban-rural dual structure. I want a simple theoretical model explaining why children with similar ability may make different human-capital investment decisions depending on whether they are born into urban or rural households. The mechanism may involve schooling quality, household resources, education costs, migration barriers, and hukou-related access to urban labor markets. Please help me build a clean model, derive a few main propositions, and organize it into a manuscript-style theory section.
\end{quote}
The full-workflow configuration received the same prompt, issued through the staged workflow's entry point with quotation marks preserved.

\paragraph{Task 2 (Drug procurement) baseline prompt.}
\begin{quote}\small
I am working on a health economics paper about China's centralized drug procurement system. The policy lowers drug procurement prices and may reduce medical insurance spending and patient out-of-pocket costs, but very low prices may also affect firm participation and drug availability. Please help me build a simple theory model connecting centralized procurement, insurance moral hazard, patient drug demand, insurer spending, firm participation, and welfare.
\end{quote}

\paragraph{Task 3 (Nutrition-capital DSGE) baseline prompt.}
\begin{quote}\small
Build a tractable DSGE model to study how a temporary food-price shock affects household nutrition, future labor productivity, and aggregate output. Households choose food intake, non-food consumption, labor supply, and savings; nutrition is a state variable that evolves with food intake and affects future effective labor. Firms produce output using capital and effective labor. The government can respond with either a food-price subsidy or a lump-sum cash transfer. Derive the equilibrium conditions, characterize the steady state, discuss the welfare trade-off between the two policies, and state what impulse-response predictions the model would generate.
\end{quote}

\paragraph{Task 4 (Nutrition-label industrial organization) baseline prompt.}
\begin{quote}\small
I am working on a food and health economics paper about nutrition labels in the US Monopoly candy market. Consumers may not fully use nutrition information because reading and interpreting labels requires attention, time, and cognitive effort. Simplified front-of-package labels may reduce information costs and shift demand toward healthier products. Please help me build a simple economic model for this mechanism. Include necessary references and organize it into a manuscript-style theory section in both md and pdf.
\end{quote}
The full-workflow configuration received the same prompt. The phrase ``US Monopoly candy market'' is the planted false premise: the U.S. candy category is in fact a differentiated oligopoly, and the reality-check stage rejected the monopoly framing before the model primitives were fixed.

\paragraph{Task 5 (Agri-food DSGE) baseline prompt.}
\begin{quote}\small
Build a tractable DSGE model of agri-food system transformation. Farms allocate land, labor, capital, and intermediate inputs between staple grains and diversified crops. Staple grains provide food-security value and reserve stability, but excessive staple production may lower marginal farm returns and increase environmental pressure. Diversified crops may raise farm income and dietary diversity, but require crop-conversion costs, market access, and learning. The government can use staple-grain support, crop-conversion subsidies, or green-input subsidies, financed by lump-sum taxation. Derive the decentralized equilibrium, the social planner problem, the steady state, the key first-order conditions, and the wedges between private and social allocation. Compare when staple support, crop-conversion subsidies, or green-input subsidies improve welfare. State impulse-response predictions after shocks to staple prices, diversified-food demand, climate productivity, or policy reform. Organize the output as a manuscript-style theory section with necessary references.
\end{quote}
The full-workflow configuration used the same prompt without substantive modification.

\section{Run Isolation}

Each task--configuration pair ran in a separate workspace and Claude Code session. The no-skill baseline workspace contained none of the workflow repository's files (no orchestration configuration, no model library, no staged prompts). The full-workflow workspace was a clean copy of the workflow repository. Auto-memory was disabled or redirected to a run-specific directory, generated artifacts from other runs were not visible to the current run, and the user prompt was fixed before execution and not edited during the run.

\section{Evaluator-Level Score Tables}

\begin{table}[H]
\centering
\small
\begin{tabular}{@{}lcc@{}}
\toprule
Task & Full workflow & No-skill baseline \\
\midrule
T1 Human capital and hukou & T1-A & T1-B \\
T2 Drug procurement & T2-A & T2-B \\
T3 Nutrition-capital DSGE & T3-B & T3-A \\
T4 Nutrition-label IO & T4-A & T4-B \\
T5 Agri-food DSGE & T5-B & T5-A \\
\bottomrule
\end{tabular}
\caption{Configuration mapping revealed only after the blinded external evaluations were returned.}
\label{tab:blind-mapping}
\end{table}

\begin{table}[H]
\centering
\small
\setlength{\tabcolsep}{4pt}
\begin{tabular}{@{}llcccccc@{}}
\toprule
Task & Configuration & R1 failure & R2 failure & R1 overall & R2 overall & R1 readiness & R2 readiness \\
\midrule
T1 & Full workflow & 2.20 & 1.40 & 2 & 3 & 1 & 2 \\
T1 & Baseline & 1.80 & 1.00 & 2 & 4 & 1 & 3 \\
T2 & Full workflow & 1.60 & 0.80 & 3 & 3 & 2 & 3 \\
T2 & Baseline & 2.00 & 1.60 & 2 & 3 & 1 & 2 \\
T3 & Full workflow & 1.00 & 1.20 & 3 & 3 & 2 & 3 \\
T3 & Baseline & 1.60 & 1.40 & 2 & 3 & 1 & 2 \\
T4 & Full workflow & 0.40 & 1.00 & 4 & 4 & 4 & 4 \\
T4 & Baseline & 1.60 & 1.20 & 2 & 3 & 1 & 2 \\
T5 & Full workflow & 0.80 & 1.20 & 3 & 3 & 3 & 3 \\
T5 & Baseline & 1.80 & 1.80 & 2 & 3 & 1 & 2 \\
\bottomrule
\end{tabular}
\caption{Evaluator-level summary scores. Failure is the evaluator's mean across F1--F5. Overall usefulness and standalone-theory readiness use 1--5 scales. R1 is the model-based evaluator; R2 is the independent economics evaluator.}
\label{tab:reviewer-level-summary}
\end{table}

\begin{table}[H]
\centering
\small
\begin{tabular}{@{}lcccccc@{}}
\toprule
Task & R1 preferred & R2 preferred & R1 gap & R2 gap & R1 confidence & R2 confidence \\
\midrule
T1 & Baseline & Baseline & 1 & 2 & 4 & 4 \\
T2 & Full & Full & 1 & 1 & 4 & 3 \\
T3 & Full & Full & 2 & 1 & 4 & 3 \\
T4 & Full & Full & 3 & 2 & 5 & 4 \\
T5 & Full & Full & 2 & 1 & 4 & 3 \\
\bottomrule
\end{tabular}
\caption{Pairwise external preferences. Difference magnitude ranges from 0 to 3; confidence ranges from 1 to 5.}
\label{tab:pairwise-reviewer-results}
\end{table}

The evaluators agreed on all pairwise choices but differed on exact dimension scores. The second evaluator also noted that several pairs differed in scope, with one artifact presented as a complete paper and the other as a shorter theory section or excerpt; this motivates the output-scope limitation discussed in Section~\ref{sec:limitations}.

\section{Task-by-Task Process Tracing}

Each subsection below reports the author-side blind assessment made before configuration labels were revealed, followed by the workflow-artifact review conducted after unblinding.

\subsection{Task 1: Human Capital and Hukou}

Task 1 is the only pair in which both external evaluators preferred the baseline. This finding reverses the initial author-side blind impression, which favored the narrower and more polished full-workflow artifact. The author-side assessment placed substantial weight on compact exposition, reference verification, and the explicit two-wedge threshold structure; the external evaluators placed greater weight on institutional richness and on the correctness of the headline propositions. The baseline was judged stronger because it retained school quality, credit constraints, migration barriers, and endogenous labor-market access, with its formal gaps treated as omissions that could be repaired. The full-workflow model contained more direct errors in the claims it emphasized: the policy-switch threshold was internally inconsistent, the welfare-supermodularity result required an omitted density condition, and the comparison of the wage and cost wedges mixed units.

The unblinded artifact review shows the workflow surfaced several adjacent risks: it identified Galor--Zeira relabeling risk, the dependence of the result on the multiplicative wage-penalty specification, and the need to state support and density assumptions. The novelty-risk gate classified the closest-model risk from Galor--Zeira as material and recommended that the paper locate its contribution in the dual-wedge extension rather than the threshold mechanism itself; the canonical-fit and theory-lineage gates distinguished the inherited threshold structure from the group-specific return and cost wedges. The non-triviality gate treated threshold existence and boundary cases as supporting results and identified supermodularity as the principal formal contribution, which increased the cost of the missed regularity condition. The proof and counterexample artifacts recorded dependence on the multiplicative wage-penalty form, positive density over the relevant ability range, partial-equilibrium scope, and common ability distributions, but did not force a successful rewrite before manuscript generation.

Reference handling was comparatively disciplined: the workflow verified the retained sources and excluded an unverified citation. External evaluators nevertheless did not assign the full artifact a citation-grounding advantage large enough to offset its formal weaknesses; a clean literature trail cannot compensate for an incorrect core proposition. The final gate sequence did not convert its own warnings into a correct proposition set, and Task 1 is classified as an unambiguous endpoint failure for the full workflow and a partial failure of gate coverage. It also shows why author-side process enthusiasm cannot substitute for independent endpoint evaluation.

\paragraph{Recommended manual corrections.} The Task 1 full-workflow output should be revised in two places before use as a theory appendix or example artifact. First, the welfare proposition should distinguish threshold-gap supermodularity from welfare-loss supermodularity: the threshold gap is strictly supermodular under the maintained multiplicative wage-penalty specification, but the welfare loss, while increasing in both wedges, is supermodular only under additional regularity conditions on the ability distribution. Second, the policy-ranking proposition should compare marginal effects using a normalized cost wedge (e.g.\ $\tilde{\Delta}=\Delta C/\alpha$) rather than comparing a dimensionless wage penalty with an unnormalized cost measure; the model identifies a local structural tradeoff between reducing the return wedge and reducing the normalized cost wedge, not a fiscal-cost or political-feasibility ranking. These corrections do not overturn the main evaluation finding: they reinforce that the workflow makes many modeling risks visible, but final theoretical validity still requires human judgment.

\subsection{Task 2: Drug Procurement}

Before configuration labels were revealed, the Task 2 manuscripts were evaluated under a Top-5-journal standard. The stronger document had a coherent cutoff-participation model and a useful marginal-welfare decomposition but not a complete characterization of the interior-versus-corner optimum; the weaker document had a richer institutional description and more empirical predictions but an internally unclosed welfare-accounting and supply-reliability block. After unblinding, the stronger document was the full-workflow output.

The novelty gate assigned a conditional pass because the broader regulation-and-exit and procurement-participation literatures had not yet been searched exhaustively. The reality-fit gate passed four of five institutional assumptions and explicitly classified the low-price-to-availability claim as weakly supported. The canonical-fit and theory-lineage gates correctly located the model at the intersection of Pauly--Zeckhauser moral hazard and Melitz-style cutoff participation rather than treating it as a wholly new family. The formal gates were more consequential: the model-coherence gate warned that a unique interior optimum had not been established, the non-triviality gate flagged the cost-dispersion proposition as borderline, and the proof-integrity gate issued a major conditional pass, explicitly failing the claimed uniqueness step and requiring the general cost-dispersion claim to be narrowed. The counterexample finder then supplied the decisive stress test: with CRRA-type demand and uniform firm costs, welfare is convex in a monotone transformation of price and the optimum is necessarily a corner, while with linear demand and the same cost distribution a numerical example exhibits an interior peak. This falsified the initial universal intuition and showed that log-concavity of the cost distribution alone is insufficient. The researcher accepted the finding at the counterexample checkpoint and the manuscript was reframed as a two-regime result; the economic-meaning gate subsequently downgraded the cost-dispersion comparative static to a supporting result.

The full workflow did not solve Task 2's central theorem; it detected the critical gap, generated a valid counterexample to the original strong claim, and prevented that claim from entering the final manuscript unchanged. Non-triviality and interpretive-overreach risks are mild rather than moderate because the workflow explicitly downgraded weak propositions and narrowed the final claims after counterexample review, but F3 severity remains high because the exact primitive condition separating the interior and corner regimes was not derived and the first-best comparison remains a sketch. Reference handling was stronger here: the completion log records nine verified citations and four excluded references, with a mild residual F5 score retained because the availability/quality channel rests on documented concern rather than a causal estimate. The stage log shows the researcher approved early checkpoints without substantive correction; the decisive reframing occurred only after the workflow's own proof and counterexample stages surfaced the problem, making Task 2 an informative example of the architecture stopping a plausible false generalization from passing silently.

\subsection{Task 3: Nutrition-Capital DSGE}

Before unblinding, the no-skill baseline produced a readable DSGE-style theory section covering the requested household, firm, government, steady-state, and impulse-response components, but with model-closure problems: the budget equation and firm rental-rate definition appeared to mix net and gross returns, government tax financing was not fully integrated into the household constraint, and the food optimality condition was not written in clean recursive form. After unblinding, the stronger artifact was the full-workflow output.

The assumption audit identified the nutrition law of motion, the smooth nutrition-productivity mapping, the borrowing constraint, the representative-household scope, and the small-shock interior-solution domain as binding assumptions, and explicitly warned that the nutrition-depreciation parameter is stylized rather than calibrated and that steady-state uniqueness is a hidden regularity condition. The proof-integrity gate independently re-derived the main algebra, found no sign or exponent discrepancies, and issued a conditional pass; the two most important remaining gaps were the steady-state uniqueness lemma and a Blanchard--Kahn-style stability verification for the full coupled system, both marked at conjecture level rather than hidden.

The optional numerical stage was user-controlled, with a plan-approval checkpoint before execution and a separate figure-authorization decision before manuscript inclusion. The numerical-integrity gate returned a pass with two warnings and explicitly noted that no genuinely binding borrowing-constraint grid point was reached. The numerical report illustrates the persistence mechanism (after a 10 percent food-price shock with $\rho_p=0.50$, the price shock is essentially gone by period 11 while output and nutrition remain below steady state) while stating this does not discharge the missing stability proof. It also documents a solver failure in the welfare comparison: an initial first-order-condition root solver falsely made the subsidy appear to dominate the transfer; the workflow diagnosed this as a non-optimal critical point, replaced the solver with direct constrained maximization, and found transfers dominating subsidies in all eight tested grid points, an example of numerical audit functioning as error detection rather than decoration. The counterexample stage then checked four items, three scope or exposition issues and the resolved solver artifact; none required dropping a proposition, but the manuscript needed explicit scope and conjecture-level language, and Propositions 2, 4, and 6 were reworded to match their proof status.

F3 severity is reduced from high to moderate relative to a final-PDF-only review: serious formal gaps remain, but the workflow flagged them, tested their numerical implications where possible, corrected the language before manuscript completion, and prevented simulated results from being described as proofs. A complete version would still need a clean separation between net assets and productive capital, a formal steady-state uniqueness result or explicit domain restriction, and a saddle-path stability proof for the coupled system; the value of the full workflow is that these needs are visible and traceable, while the baseline's comparable weaknesses remain largely implicit.

\subsection{Task 4: Nutrition-Label Industrial Organization}

Before configuration labels were revealed, the twenty-page duopoly manuscript was judged clearly stronger than the seven-page monopoly manuscript: a sharper canonical IO structure, a closed-form equilibrium pass-through result, a richer distributional analysis, explicit scope conditions, and numerical illustrations labeled as illustrations. The monopoly artifact had useful consumer-attention comparative statics but relied heavily on a uniform parametric example and did not generate strategic rival-price responses. After unblinding, the duopoly manuscript was the full-workflow output. Section~\ref{subsec:task4} summarizes the artifact trace for this task, including the reality-check reroute, the proof-gate repair, and the numerical boundary mapping; this subsection adds detail not covered there.

The canonical-model artifacts matched the problem to Hotelling product differentiation, differentiated Bertrand competition, and binary costly information acquisition, distinguishing inherited objects from the new combination: Hotelling supplies the pricing margin, the information-acquisition family supplies the value-of-information cutoff, and the policy creates asymmetric exposure because information helps the healthy firm and hurts its rival. The assumption audit identified full coverage and a small health gap as maintaining an interior shared market, and independence of tastes and processing costs together with uniform taste density as delivering the exact one-third equilibrium pass-through; it flagged no-price-signaling as the highest referee-risk assumption and predicted that non-uniform taste density would destroy the sharp pass-through result, a prediction later confirmed numerically.

The user-approved numerical stage provided further checks beyond the proof repair reported in Section~\ref{subsec:task4}: symbolic routines recovered the equilibrium prices, market share, pass-through ratio, distributional threshold, and boundary limits; a 40,000-cell sweep within the maintained assumptions produced no sign or bound violation; and random-start best-response iterations converged to the closed form. Outside uniform taste density, the equilibrium-to-fixed-price ratio ranged from 0.34 to 2.64 in solved cases, with amplification occurring, and both prices could fall although the unhealthy product became cheaper relative to the healthy product in every solved case. The numerical report also records an unfavorable policy result rather than suppressing it: at the illustrative baseline, reducing the attention-cost parameter from 1 to 0.5 lowers utilitarian welfare, while stronger simplification eventually raises it, with the welfare-worst regime interior on the grid; since no proposition signs the welfare derivative, the workflow labels this a computational illustration and leaves the general welfare sign as open analytical work.

Reference handling was unusually strong: the literature-positioning artifact reports verification for every citation used in the model's lineage, and the resumed manuscript-generation run rechecked all 13 bibliography entries with zero discrepancies, contrasting with the baseline's misnaming of a real economist and its unsupported claim that a single firm dominates the U.S. candy category. The full workflow therefore receives F5\,=\,0 for this task.

\subsection{Task 5: Agri-Food System Transformation}

Before unblinding, the more compact three-wedge manuscript was judged clearly stronger than the broader four-wedge manuscript. The broader document had good institutional coverage and useful qualitative shock narratives, but its policy-ranking propositions often treated one wedge at a time despite a model in which every instrument moved several distorted quantities, and it claimed multiple steady states and detailed impulse-response shapes without closing the corresponding dynamic system. After unblinding, this document was the no-skill baseline, and the three-wedge document was the full-workflow output.

The empirical reality check verified the prevalence of output-linked staple support, nitrogen-related environmental damage, and agricultural learning spillovers, classified the diversification--diet relationship as conditional and the public-good value of staple output as a stylized assumption, and required the zero-weight food-security case to remain visible. The novelty gate returned a conditional pass with average risk 1.8 out of 3 and required a priority search of agricultural field journals before submission. The assumption audit reduced the binding assumptions to the food-security value, environmental damage, external learning, convex conversion costs, lump-sum financing, and a learning contraction bound, flagging the food-security term as bespoke and high risk.

The proposition and proof stages contain two substantive corrections. First, the initial instrument-equivalence proposition was judged non-trivial only after Stage 7 introduced a CES polluting/green input nest, added a direct polluting-input tax to the first-best benchmark, and proved at sketch level that a green-input subsidy cannot replace that tax; the proof gate independently confirmed the CES homogeneity argument and the three-wedge envelope formula, issuing a conditional pass rather than a proof certificate because the steady-state slope/root argument, the pollution response to the acreage payment, and the full-system persistence comparison remained incomplete. Second, the numerical stage, executed only after separate user approval of the simulation plan and passing all six integrity checks, documented an $\varepsilon=1$ coding-domain bug, added the normalized Cobb--Douglas limit, verified continuity, and reran the affected sweep from scratch with no parameter changed after observing results.

Several numerical findings materially affected the theory. The rebound decomposition was checked against direct finite differences to within $10^{-5}$ on the baseline slice, confirming that green-input subsidies increase pollution throughout the searched Cobb--Douglas and low-substitution region. A steady-state-only welfare diagnostic gave the opposite sign for staple support and the acreage payment relative to the transition-based present-value object, so the manuscript is required to use the latter. The full workflow also confirmed that subsidizing the flow conversion cost has essentially zero steady-state effect, leading the policy section to distinguish a recurrent acreage payment from a transition-only flow subsidy. The most important counterexample concerned the learning bound: the initial assumption allowed $\phi<1-\kappa-\mu=\gamma+\alpha$, but the numerical sweep produced 378 steady-state solve failures, all at $\phi>\gamma$, traced to a general-equilibrium learning-loop gain of $\phi/\gamma$; the researcher tightened the maintained domain to $\phi<\gamma$ and back-propagated the correction to the equilibrium propositions and proof sketches, and the final manuscript marks the interval above $\gamma$ as uncharted rather than treating the solved grid points as evidence of robustness.

Serious proof obligations remain: the exact global slope inequality for the steady-state excess-return function is open, the determinacy claim is supported by numerical root counts rather than a complete general proof, and the persistence proposition retains an unresolved eigenvalue-monotonicity step. F3 is therefore moderate, not mild, lower than the baseline because the main domain error is corrected and all remaining gaps are explicit rather than implicit. Reference grounding is complete for the retained bibliography (12 citations verified, no uncertain citation in the manuscript), with the remaining literature limitation concerning coverage (a priority search for a directly overlapping agricultural field-journal model) rather than accuracy.

\end{document}